# Can large (M ≥ 8) EQs be triggered by tidal (M1) waves? An analysis of the global seismicity that occurred during 1901 – 2011.

Thanassoulas[1], C., Klentos[2], V., Verveniotis, G.[3], Zymaris, N.[4]

1. Retired from the Institute for Geology and Mineral Exploration (IGME), Geophysical Department, Athens, Greece.
   e-mail: thandin@otenet.gr - URL: www.earthquakeprediction.gr

2. Athens Water Supply & Sewerage Company (EYDAP),
   e-mail: klenvas@mycosmos.gr - URL: www.earthquakeprediction.gr

3. Ass. Director, Physics Teacher at 2nd Senior High School of Pyrgos, Greece.
   e-mail: gver36@otenet.gr - URL: www.earthquakeprediction.gr

4. Retired, Electronic Engineer.

**Abstract**

The analysis of two global data sets of large earthquakes (2010 - 2011, 30 samples of M ≥ 7R and 1901-2011, 178 samples of M ≥ 8R) reveals that there exists a cause and effect relation between the vertical tidal **M1** component amplitude peak and the time of occurrence of the latter EQs. A physical model mechanism is postulated that justifies the obtained results. It is shown that the tidal waves can trigger a large EQ, despite their small amplitude, provided that the seismogenic area is under critical stress load conditions. Actually, it is shown that a large EQ can be triggered by the cooperative action of all vertical tidal components but mainly by the **M1** and **K1** ones. Examples are presented from the most recent global large EQs (Summatra, Mw = 9.1, 2004 and Japan, Mw = 9.0, 2011) and from Greece (Kythira, Greece, Ms = 6.9R, 2006 and Skyros, Greece, Ms = 6.1R, 2001). The postulated physical model provides the means for the implementation of the first step towards a really short-term earthquake prediction.

## 1. Introduction.

The Earth-tides mechanism was early recognized as a potential trigger for the occurrence of strong earthquakes. This approach was followed to study the time of occurrence of EQs and their correlation to Earth-tides. Knopoff (1964), Shlien (1972), Heaton (1982) and Shirley (1988) suggested the Earth-tides as a triggering mechanism of strong EQs, Yamazaki (1965, 1967), Rikitake et al. (1967) studied the oscillatory behaviour of strained rocks, due to Earth-tides, while Ryabl et al. (1968), Mohler (1980) and Souriau et al. (1982) correlated Earth-tides to local micro-earthquakes and aftershock sequences. Emter (1997) studied the tidal triggering of earthquakes and volcanic events, Glasby and Kasahara (2001) studied the influence of tidal effects on the periodicity of earthquake activity in diverse geological settings.

In order to physically investigate the relation between the Earth tide and fault rupture, recent studies have examined stress components resolved on the fault plane by using the focal mechanism solution (Tsuruoka et al. 1995; Vidale et al. 1998; Wilcock, 2001; Tanaka et al. 2002a, b). Positive tidal – earthquakes correlation was reported too by Heaton (1975), Young and Zurn (1979), Souriau et al. (1982), Ulbrich et al. (1987), Rydelek et al. (1988), Tolstoy et al. (2002), Crockett et al. (2006)

Other studies suggested that the solid Earth tides do not trigger earthquakes (Cotton, 1922; Klein, 1976; Heaton, 1982; Lockner and Beeler, 1999; Cochran et al. 2007) although the produced by the tidal forces varying stresses in the Earth's crust present a daily maximum shear stressing rate that is 2 orders of magnitude larger than the rate of accumulation of stress along active faults due to plate motion (Heaton, 1982). Careful analyses of quality data sets suggest little or no correlation between Earth tides and earthquakes (e.g., Vidale et al. 1998). The same conclusions were reached by Beeler and Lockner (2003) who suggested that a rather large > 13000 number of EQs is required in order to detect the very weak correlation of earthquakes to the daily Earth tides. On the basis of early experiments on static fatigue Knopoff (1964) suggested that stress increases equivalent to the tidal stress amplitude would induce failure following a delay of months or even years. He concluded that earthquakes would not correlate with periodic stressing of such small amplitude. Similarly, Rydelek et al. (1992) concluded that the lack of correlation between earthquake occurrence and the tides indicates that failure is intrinsically time-dependent with a characteristic time greater than the tidal period.

A characteristic feature of most of the studies that concern tidal waves correlation to seismicity is that deal mostly with daily tidal. In a very limited number the "biweekly" tidal variation is considered (i.e. Crockett et al. 2006; Sue, 2009). At this point it is interesting to present the corresponding objections paragraph of Cochran et al. (2007) regarding this topic.

"However, none of the studies report seeing a correlation with the biweekly tide. Even regional studies of far larger sets of earthquakes in the San Francisco Bay Area (McNutt and Heaton, 1981; Kennedy et al. 2004) and the Pacific Northwest (Kennedy et al. 2004) find no measurable correlation. As stated by Hartzell and Heaton (1989), the difference in the amplitude of the biweekly tidal stress envelope is much smaller than the diurnal peak-to-peak stress variation, with diurnal stress variation being over 5 times larger than the biweekly variation. Along the same vein, theoretical studies predict a correlation with the diurnal but not biweekly tides due to the small overall amplitude variation of biweekly tides (Dieterich, 1987)".

Some more characteristic features of earlier studies are the use of the correlation of the Lunar "phases" to the related seismicity while the used tidal forces were projected along the fault plane. Laboratory experiments has shown that peak stress occurs prior to failure while there is a time lag between the onset of a detectable slip and the time of failure defining a measure of the duration of the nucleation.

A quite different approach, in order to investigate the relation of the tidal waves to the time of triggering large EQs, was followed by Thanassoulas (2007). In the specific analysis the excess stress load required to trigger an EQ is attributed to the local lithospheric deformation at the regional place where the tidal force is temporarily applied. Therefore, the imminent large EQ is triggered not directly by stress increase due to the tidal waves, but indirectly by the corresponding lithospheric deformation as it will be shown in the following presentation.

Since the tidal forces are a global physical phenomenon, the triggering of large EQs must be a global one and must be observed all over the world regardless specific local geology and EQ tectonic generating mechanisms. Therefore, the aim of this study is to perform a global analysis of the time of occurrence of large EQs in terms of the **M1** local tidal component and to identify the existing correlation if any.

## 2. Theoretical analysis

The creation of fracture zones and faulting of the Earth is due to the present stress field conditions which are applied in a particular area. Fracturing takes place in various modes (Mattauer, 1973), when extensional or compressional forces are applied on it. In the following figure (1) the black arrows indicate the applied forces, while the lines in the solid block indicate the various generated modes of fracturing. In case (**A**) there is block movement which creates a typical thrust fault combined with internal micro-fracturing typically normal and parallel to the direction of the applied stress field. In case (**B**) the generated, internal, micro-fracturing is interconnected and delineated diagonally in respect to the applied stress field direction, while in case (**C**), a characteristic strike-slip fault has been formed, due to exceptional type (**B**) stress field conditions.

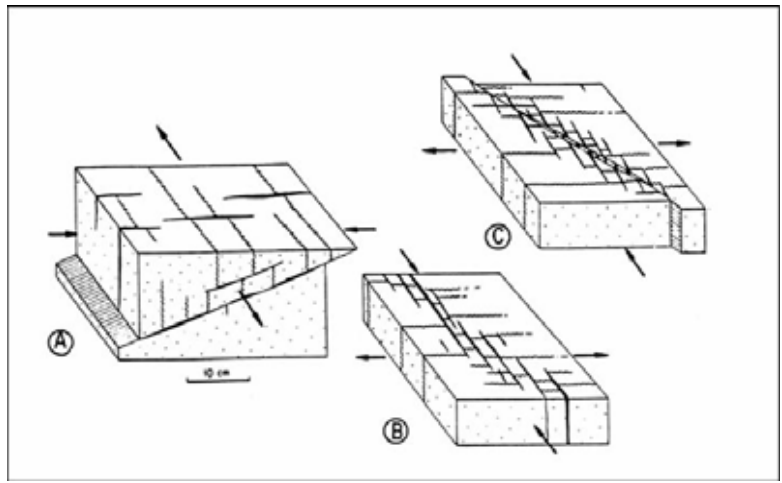

Fig. 1. Fracturing – faulting which is generated by applied stress (Mattauer, 1973). The arrows indicate the applied stress direction (compressional, extensional, shear-stress).

The tectonic stress load that is applied at a seismogenic area by the pairs of forces shown in figure (1) increases or decreases according to the excess load added by the lithospheric deformation due to its oscillation that is triggered by the tidal waves. The deformation of the lithosphere follows the oscillatory character of the Earth-tides. Garland (1971), Stacey (1969), Sazhina and Grushinsky (1971), study in detail this type of Earth's oscillation, not to mention the majority of the Geophysical textbooks. The lithosphere, following the Earth-tide oscillatory forces, self-oscillates with maximum amplitude which varies as follows:

**According to Garland (1971),** the maximum elevation of the lithosphere is of the order of 10cm

**According to Sazhina and Grushinsky (1971),** the lithosphere oscillates with a maximum p – p value of 53.4cm for the Moon component, while for the Sun component is only 24.6cm.

That means that, the maximum oscillatory amplitude can reach the value of 78cm p – p, in cases, when the two components are added in phase.
A sketch drawing of the tidal forces applied on the Earth's surface id shown in the following figure (2).

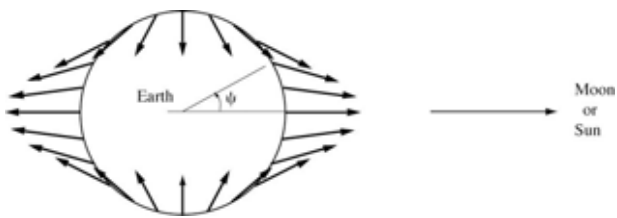

Fig. 2. Tidal force vectors at earth's surface at $15^0$ increments of the zenith angle **ψ** (Hale, 2009).

Let us consider a tidal force (**T**) applied on a certain place of the earth's surface shown in figure (3) left. That tidal force can be decomposed as: a normal to the ground surface tidal component and as components along the fault strike direction and normal to the fault strike direction (fig. 3).

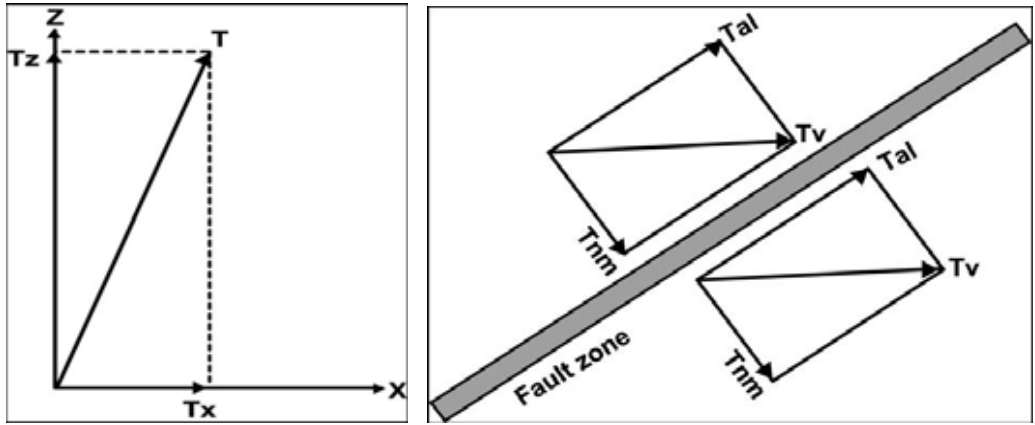

**Fig. 3.** Left = Tidal force (**T**) applied on earth's surface and decomposed into a horizontal **Tx** and a vertical **Tz** component. Right = horizontal component (**Tv**) decomposed into: along the fault direction (**Tal**) and normal to the fault direction (**Tnm**) components.

If we compare the tidal components of figure (**3**) to the compressional or extensional pairs of forces of figure (**1**) that generate a specific type of faulting it is clear that the tidal forces cannot generate / activate any type of fault due to the fact that they do not form a compressional or extensional pair of forces. Moreover, they cannot directly trigger any EQ since the applied tidal effect is the same at the adjacent block sides of the specific fault and therefore no differential movement (EQ) can be generated between the two adjacent fault blocks. Thus, the question that arises is how the tidal forces increase or decrease the stress load of a seismogenic area? The vertical component of the tidal forces applied at any place on the earth's surface is of particular interest in answering the posed question.

Firstly let us consider the case of two tidal forces T1 and T2 while T1< T2 in magnitude (**fig. 4** left). The corresponding vertical components are related as:

$$T1v > T2v \text{ or } T1\sin(i) > T2\sin(j) \qquad (1)$$

Equation (**1**) is valid for certain ranges of **i, j** and **T1, T2** amplitudes thus suggesting that a smaller tidal vector can, under certain conditions, to apply a stronger effect upon the lithosphere than a larger tidal vector. The lithosphere, due to the vertical tidal component is forced to deform and oscillate as shown in figure (**4**) right.

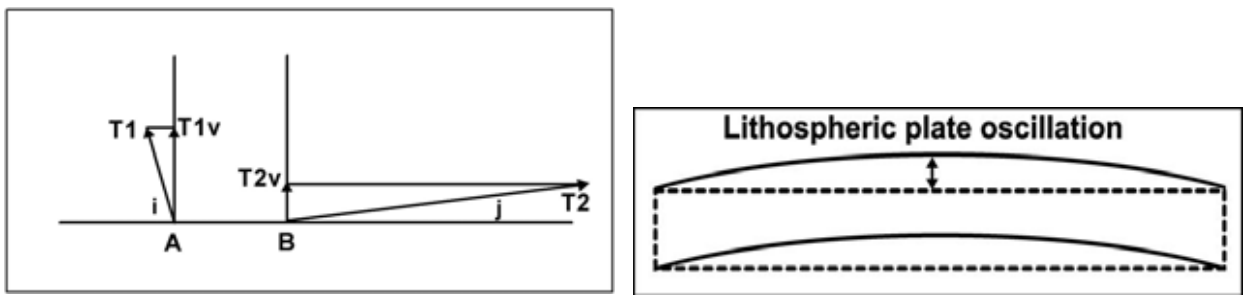

**Fig. 4.** Tidal forces decomposed into its vertical component (left) and the corresponding triggered lithospheric deformation – oscillation (right).

Therefore, the comparison of the earthquake occurrence time to the maxima of the tidal forces (generally) may lead to erroneous results concerning their cause and effect relation. On the contrary, the vertical component is the driving mechanism of the lithospheric oscillation. Consequently, a close relation of the oscillating stress load of the lithosphere and the vertical tidal component is most appropriate to exist.

In the following figures (**5, 6**) is demonstrated the mechanism that generates increase / decrease of the stress load in a seismogenic area due to the vertical tidal component.

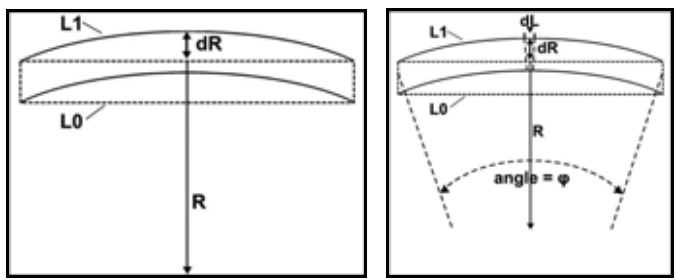

**Fig. 5.** Vertical tidal forces increase earth's radius at **dR** (left) and cause (right) increase of the lithosphere length for **dL** (considering 1- D deformation).

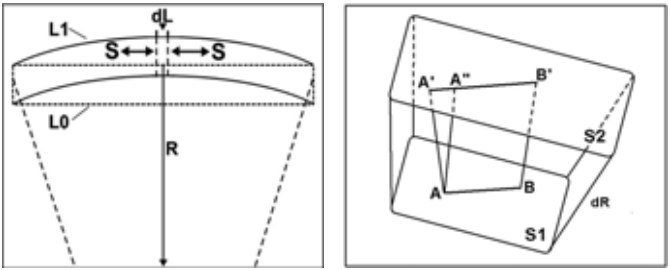

Fig. 6. The lithosphere is expanded at **dL** and therefore an oscillating stress load (**S**) is generated (left). Right = horizontal deformation (**A' – A''**) observed between two arbitrary selected points **A** and **B** of the lithosphere.

In figure (6, right) a surface **S1** is deformed (expanded) into **S2** due to tidal oscillation. Therefore, any, arbitrary in direction, line segment A - B is deformed into A' - B' and consequently the observed strain is:

$$\text{Observed strain } S = A' - A'' / A-B \qquad (2)$$

The latter implies directly that the tidal stress load that is induced at any place on the earth's surface is of omni-directional (2-D) character. In other words that means that it is independent from the direction and further more it is always added on the tectonic stress present in any seismogenic area, no matter what is the fault strike direction that will be activated in the future. The latter is quite different from what Sue (2009) stated as "earth tides work as a trigger of earthquakes when the direction of compressional stress caused by the earth tide coincides with that of the P-axis of a focal mechanism at a fault".

An argument against tidal triggering of large earthquakes by the "biweekly" (it means the **M1** component) tidal variation is that its amplitude is much smaller (Dieterich, 1987; Hartzell and Heaton, 1989; Cochran et al. 2007) than that of the diurnal (daily) variation (**K1**). The following postulated model for earthquake triggering by tidal waves justifies the triggering effect of the "biweekly – M1) and simultaneously earthquake triggering by the diurnal- daily (**K1**) tidal component.

It is assumed that at a certain seismogenic area are simultaneously present a) the tectonic stress of low rate, b) the low amplitude "biweekly – M1" oscillating tidal stress component, and c) the large amplitude (compared to the M1) diurnal (K1) tidal oscillating stress variation. All these stress loads act simultaneously superposed on each other on the seismogenic area as it is shown in the following figure (7). Firstly, the tectonic stress is represented by the low rate green line. Next, on top of it is superposed the large amplitude diurnal **K1** daily variation stress load represented by the black line. Finally, the **M1** "biweekly" oscillating stress load is represented by the envelope of the daily diurnal variation. The horizontal red line represents the rock failure stress load level. In order to trigger an earthquake the only requirement is that the sum, at a certain time, of all stress loads equals the rock fracture stress level (red line). Let us examine two distinct areas of figure (7). At area (**A**) the total stress load is smaller than what is required for rock fracturing (EQ). At area (**B**) the total stress load equals the rock fracturing level. At place (**B**) an earthquake will be triggered. It is clear from figure (7) that the triggered EQ will occur on the peak of the M1 and the daily (K1) diurnal tidal variation.

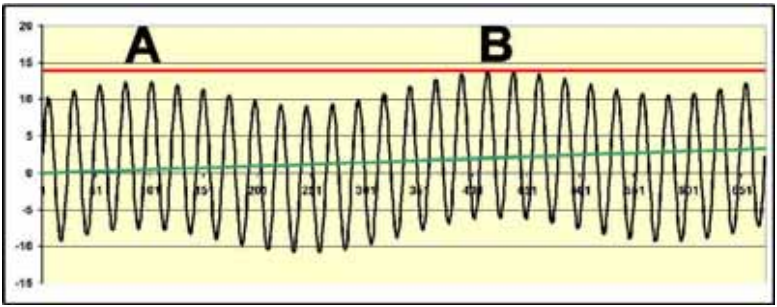

Fig. 7. Mechanism that explains how the tectonic stress (green line), the daily (**K1**) diurnal tidal tectonic stress and the "biweekly – **M1**" stress load interact in a seismogenic area in order to trigger an earthquake. Area **A** = long before triggering. Area **B** = triggering of earthquake.

Following are some indicative examples from real recent large EQs. The vertical tidal component has been calculated, by the Rudman et al. (1977) method, at a day's time interval for each large EQ at its geographical coordinates.

**The large EQ (Mw = 9.0) of Japan of the 11$^{th}$ of March, 2011.**

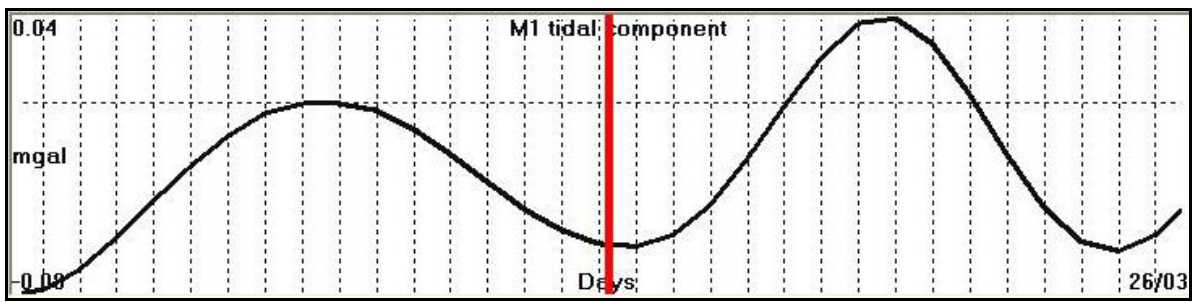

Fig. 8. Comparison of the **M1** tidal oscillation (black line, the lithosphere is forced to oscillate in the same mode) with the time of occurrence (red bar) of the Japan EQ of March 11th, 2011 (**Mw = 9.0**). Vertical scale is in mgals.

**The large EQ (Mw = 9.1) of Sumatra of the 26th of December, 2004.**

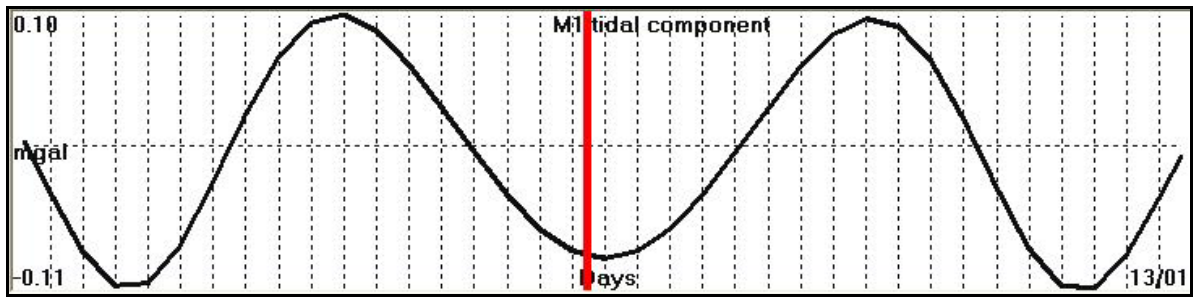

Fig. 9. Comparison of the M1 tidal oscillation (black line, the lithosphere is forced to oscillate in the same mode) with the time of occurrence (red bar) of the Sumatra EQ of December 26th, 2004 (**Mw = 9.1**). Vertical scale is in mgals.

Both EQs occurred on the peak (+/- a few hours) of the **M1** component. The latter analysis will be applied on global data for large EQs as follows.

**3. Data description**

Two different data sets were used in this study.

The first one is the **EMSC** earthquake catalog for 2010 – 2011. This catalog was used for a preliminary global test of the previous analysis for a rather short period of time and for earthquakes with magnitude **M ≥ 7R**. The No. of EQ samples used is 30.

The second data set is the **USGS** catalogs for the period 1901 – 2011. The study period was limited to 1901 due to the fact that the tidal generating program has an early limit concerning the Julian dates. The No. of EQ samples is 178 with magnitude **M ≥ 8R**.

**4. Results**

For both the data sets the deviation of each earthquake occurrence time from the corresponding **M1** tidal peak was calculated. The results were used to construct diagrams for the EQ occurrence frequency vs. deviation in days from the **M1** tidal peak. These diagrams are presented in the following figures (**10, 11**).

**a. 2010 - 2011 data for M ≥ 7R (30 data samples)**

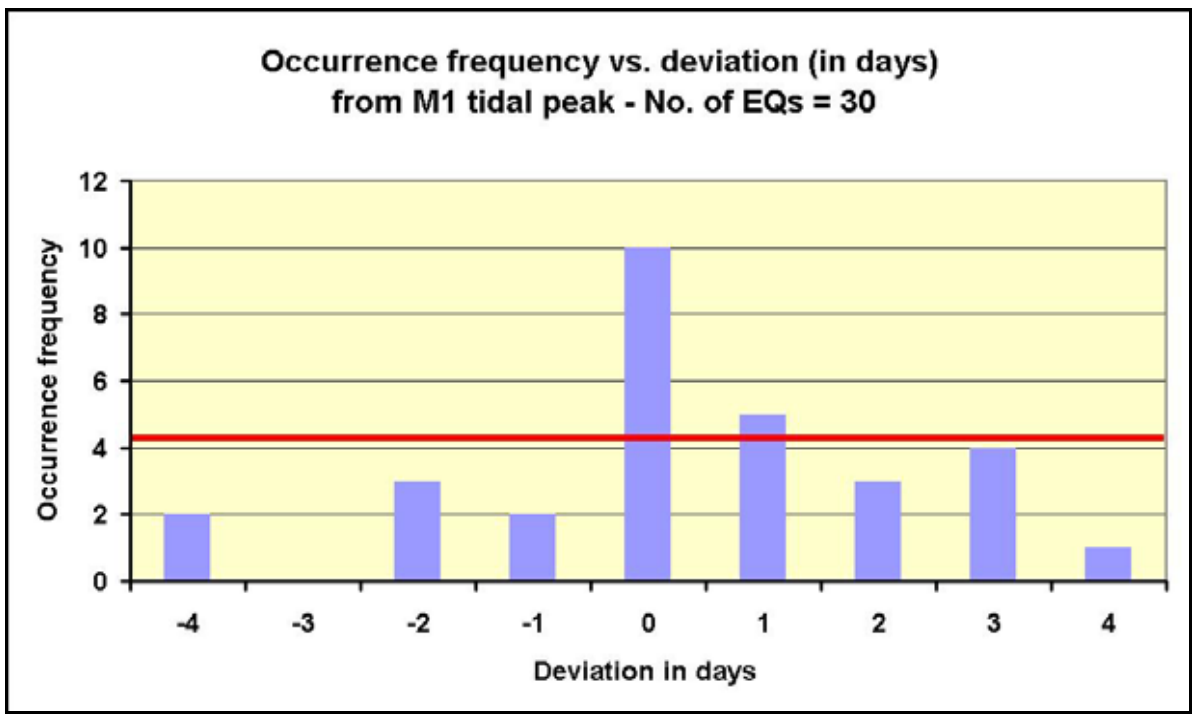

Fig. 10. Occurrence frequency vs. deviation (in days) from **M1** tidal peak. Study period: 2010 – 2011, **M ≥ 7R**. Global data from **EMSC**

The "by chance" probability $P_w$ of an earthquake to deviate for a certain no. of days within a half period (7 days) of **M1** is:

$$P_w = 1 / 7 = 14.28\% \quad (3)$$

Therefore, the "by chance" no. of EQs $P_{ds}$ that will deviate for a certain no. of days, for the present data set, is:

$$P_{ds} = 14.28\% \times 30 = 4.286 \quad (4)$$

The $P_{ds}$ value is represented in figure (**10**) by a red horizontal line. What ever seismic events are below the red line can be considered as random ones while all above it can be considered as triggered by the **M1** tidal component. A characteristic peak is observed at the zero deviation value. The ratio (**R**) of the No. of EQs, above the random threshold with zero deviation to the no. of EQs for a random threshold (4.286) indicates how strong the influence of **M1** in triggering large EQs is.

$$R = (10 - 4.286) / 4.286 = 1.3331 \text{ or } 133.31\% \quad (5)$$

**b. 1901 – 2011 data for M ≥ 8R (178 data samples)**

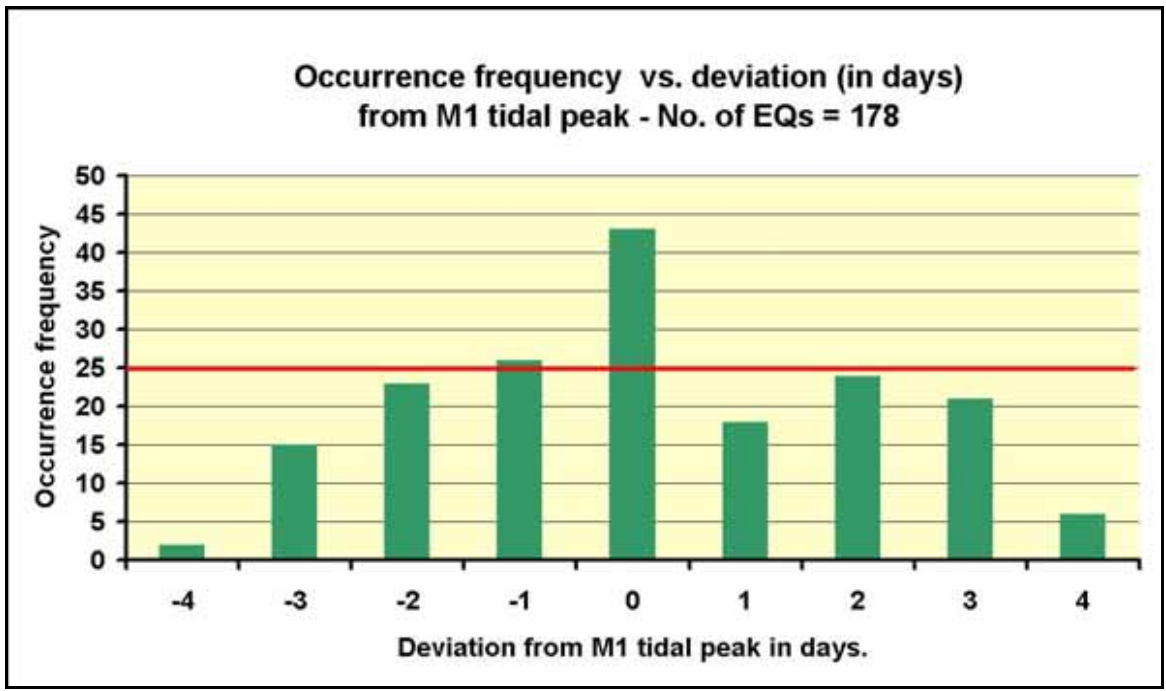

Fig. 11. Occurrence frequency vs. deviation (in days) from **M1** tidal peak. Study period: 1901 – 2011, M ≥ 8R. Global data from USGS

The "by chance" no. of EQs with a $P_{ds}$ value that will deviate for a certain no. of days, for the present data set, is:

$$P_{ds} = 14.28\% \times 178 = 25 \quad (6)$$

And

$$R = (43 - 25) / 25 = 0.72 \text{ or } 72\% \quad (7)$$

In both cases it is clear that the **M1** tidal component is closely related to the most significant mechanism that can trigger large EQs.

**5. Discussion - Conclusions.**

The majority of the reports that deal with the triggering of earthquakes by the tidal waves usually refer to a large number of earthquakes and of variable magnitude from small seismicity to the large one. Although this approach presents certainly a scientific interest, it is far more interesting to study this relation having in mind the possibility of a short-term earthquake prediction. The occurrence of large EQs is not only an outstanding nature's phenomenon but it can also create large damages in society and kill people too. Therefore, in this study as data sets are taken only large EQs, some of them being very disastrous.

Although in most reports the "biweekly" or **M1** tidal component has been rejected as a possible EQ tidal triggering mechanism, the postulated physical mechanism predicts and justifies in terms of physics the term "when" a large EQ may occur, provided that the seismogenic area is at a critical state of stress load. Following this model two tests were performed: a) a short one on a global data set of 30 EQs (2010 - 2011, M ≥ 7R) and a second one on a global data set of 178 EQs (1901- 2011) of M ≥ 8R of variable tectonic setting and focal mechanisms both.

The results of this analysis have shown that the vertical component of **M1** triggers a significant number of large EQs exactly on the peak of its amplitude. This is due to the fact that the lithosphere is driven at maxima of stress load at the same time as the **M1** amplitude peak. The no. of the EQs that coincide with the **M1** peak is well above (**133.31%** and **72%**) from what is expected as a random coincidence. Very similar results were obtained by Thanassoulas (2007) from the analysis of the large EQs that occurred in the Greek territory.

The postulated physical mechanism of figure (7) implies a very interesting outcome. When the seismogenic area has reached an **M1** peak (and is under critical stress load conditions) then the last stress load bit required to trigger the EQ will be added by the daily – diurnal **K1** variation. Two typical such examples are presented from the recent years Greek seismicity. The first one (**fig. 12**) refers to the 20060108 Kythira, Greece EQ (Ms = 6.9R).

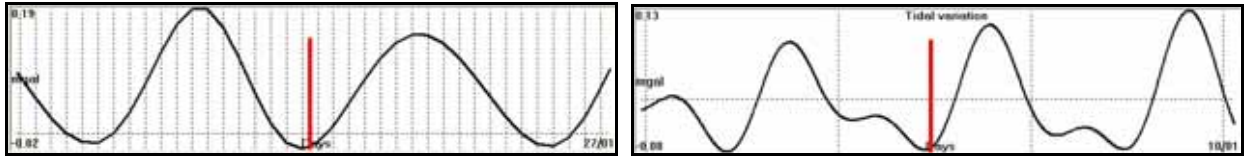

Fig. 12. **M1** tidal variation (left) and **K1** diurnal (right) for the 20060108 Kythira, Greece EQ (**Ms = 6.9R**). The red bar indicates the EQ occurrence time.

The Kythira EQ was delayed for **43** minutes after the stress load maxima.

The second example (**fig. 13**) refers to the Skyros, Greece EQ, 26/7/2001, **Ms = 6.1R**.

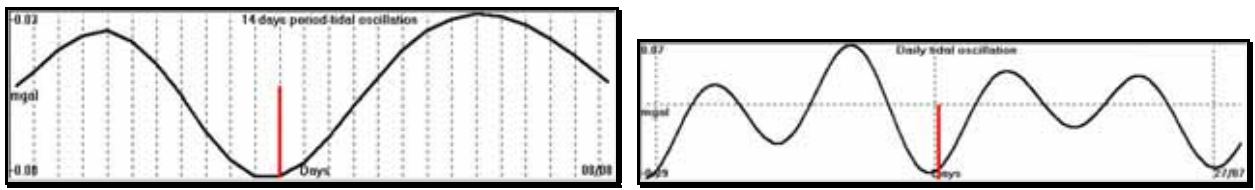

Fig. 13. **M1** tidal variation (left) and **K1** diurnal (right) for the the EQ in Skyros, Greece, 26/7/2001, **Ms = 6.1R**.

The Skyros EQ was delayed for **41** minutes after the stress load maxima.

Assuming that the time of a detectable slip ($t_{onset}$) coincides with the time of the stress peak ($t_{peak}$) and setting as ($t_{fail}$) the time of the EQ occurrence, then the nucleation time ($t_n$) can be calculated (Beeler et al. 2003) as:

$$t_n = t_{fail} - t_{onset} \quad (8)$$

That is:   $t_n$ = 43 minutes for the Kythira EQ
          $t_n$ = 41 minutes for the Skyros EQ

For both cases the nucleation time is much more smaller than the period of the triggering tidal waves of **M1** (14 days) and of the daily diurnal **K1** (24hours), therefore they comply to the requirements set by Beeler et al. (2003) for accepting a tidally triggering mechanism of an EQ. It is clear that the negative conclusions concerning tidal (biweekly – **M1**) triggering of large EQs, are not valid when applied particularly on large EQs.

It is more interesting that these positive results of the present study were obtained without taking into account the local tectonic conditions or focal mechanisms of each analysed EQ. The postulated physical model is a universal one and the tidal forces are too. Therefore, since there is a cause and effect relation between them it was expected that a global application of the model would provide positive results as obtained in this study.

Finally, the presented model is the first step for a really short-term time prediction of a large EQ. The practical use of this model has been presented in detail by Thanassoulas (2007) and Thanassoulas et al. (2010).

## 6. References.